\documentclass[twocolumn,showpacs,amsmath,amssymb]{revtex4}
 
\usepackage{graphicx}
\usepackage{dcolumn}
\usepackage{bm}

\begin{document}
\newcommand{\be}{\begin{equation}}
\newcommand{\ee}{\end{equation}}
\newcommand{\bea}{\begin{eqnarray}}
\newcommand{\n}{\nonumber\\}
\newcommand{\eea}{\end{eqnarray}}

\title{\bf Centrality Measures in Spatial Networks of Urban Streets}

\author{Paolo Crucitti$^{1}$, Vito Latora$^{2}$ and Sergio Porta$^{3}$}

\affiliation{$^{1}$  Scuola Superiore di Catania, Italy}
\affiliation{$^{2}$  Dipartimento di Fisica e Astronomia, Universit\`a di Catania, 
and INFN Sezione di Catania, Italy}
\affiliation{$^{3}$  Dipartimento di Progettazione dell'Architettura, 
Politecnico di Milano, Italy} 
\date{\today}

\begin{abstract}
We study centrality in urban street patterns of different world cities represented as 
networks in geographical space. 
The results indicate that a spatial analysis based on a set of four centrality 
indices allows an extended visualization and characterization of the city structure.  
Planned and self-organized cities clearly belong to two different universality classes. 
In particular, self-organized cities exhibit scale-free properties similar to those 
found in the degree distributions of non-spatial networks. 
\end{abstract}


\vspace{0.5cm}
\maketitle

A particular class of complex networks \cite{report} are those embedded in the real space, 
i.e. networks whose nodes occupy a precise position in two or three-dimensional 
Euclidean space, and whose edges are real physical connections. 
With a few exceptions \cite{yook02,ants04,newman04}, 
most of the works in the literature have focused on the characterization of the 
topological (relational) properties of {\em spatial networks}, 
while the spatial aspect has received less attention, when not neglected at all. 
However, it is not surprising that the topology of spatial networks is
strongly constrained by their geographical embedding. 
For instance, the number of long range connections \cite{yook02,ants04,newman04} 
and the number of edges that can be connected to a single node \cite{ants04},  
is limited by the spatial embedding. 
 This is particularly evident in planar networks (e.g. those networks
forming vertices whenever two edges cross) \cite{west95}, as urban streets or 
ant networks of galleries \cite{ants04}, and has important consequences 
on the possibility to observe a small-world behavior or scale-free degree 
distributions \cite{report}. 
Consequently, spatial networks are different from relational networks, 
and as such they need to be treated. 

{\em Centrality} has remained a fundamental concept in network analysis 
since its introduction in structural sociology \cite{bavelas48,wasserman94}.  
The network approach has also a long tradition in economic geography 
and city planning, where it has been used to investigate the 
territorial relationships among communication flows, population, wealth 
and land-uses \cite{wilson}.  
However, when dealing with {\em urban street patterns}, centrality has been 
studied in relational networks only \cite{hillier1,jang,porta_epb1}, 
neglecting a fundamental aspect as the geography. 
In such an approach, known as the {\em dual representation} \cite{jang,porta_epb1} or 
{\em information city network} \cite{rosvall}, a city is transformed into a relational 
(topological) graph by mapping the streets into the graph nodes and the 
intersections between streets into edges between the nodes. 
In the present Letter, we study centrality in urban street patterns of 
different world cities represented as spatial networks.  
In our approach, that is opposite to the dual one, we work within a fully metric  
framework in which the distance has to be measured not just in topological 
terms (steps), like in the dual representation of a city \cite{jang,porta_epb1,rosvall} 
or in social \cite{wasserman94} and other complex systems 
\cite{report}, but rather in properly spatial terms (meters, miles). 
The results indicate that a spatial analysis based on a set of different 
centrality measures (properly extended for spatial graphs) allows: 
1) a visual characterization of the structural properties of a city; 
2) the evidence that planned and self-organized cities belong to two 
different universality classes;   
3) to find scale-free properties similar to those found in the degree 
distributions of relational (non-spatial) networks.

%
We have selected eighteen 1-square mile samples of different world cities 
from Ref.~\cite{jacobs}, 
imported them in a GIS (Geographic Information System) 
environment and constructed spatial graphs of street networks.  
In our approach, each urban street sample is turned into a undirected, valued graph $G$,   
where intersections are nodes and streets are edges. We denote by $N$ the 
number of nodes and by $K$ the number of edges. 
The nodes are characterized by their position $\{ x_i,y_i \}_{i=1,...,N}$ in the unit square. 
The obtained graphs can be described by the adjacency matrix $A$, 
whose entry $a_{ij}$ is equal to 1 when there is an edge 
between $i$ and $j$ and 0 otherwise, and by a matrix $L$,
whose entry $l_{ij}$ is the value associated to the edge, in our
case the length of the street connecting $i$ and $j$. 
\begin{table}
\caption{Basic properties of the spatial graphs obtained from eighteen 1-square 
mile samples of different world cities. 
\label{table1}} 
\begin{tabular}{l|l|l|l|l|l|l|l|l}
  & CASE         &   N  &  K   &~~ &    & CASE  &   N       &  K        \\
\tableline
1 & Ahmedabad    & 2870 & 4387 &~~ & 10 &Paris  &   335     &   494     \\
2 & Barcelona    & 210  & 323  &~~ & 11 & Richmond & 697 & 1086\\
3 & Bologna      & 541  & 773  & & 12 & Savannah & 584 & 958      \\
4 & Brasilia     & 179  & 230  & & 13 & Seoul    & 869 & 1307     \\
5 & Cairo        & 1496 & 2255 & & 14 & San Francisco & 169 & 271 \\
6 & Los Angeles  & 240  & 340  & & 15 & Venice   & 1840 & 2407    \\
7 & London       & 488  & 730  & & 16 & Vienna   & 467  & 692     \\
8 & New Delhi    & 252  & 334  & & 17 & Washington   &192 & 303   \\
9 & New York     & 248  & 419  & & 18 & Walnut Creek &169 & 197   \\
\end{tabular}
\end{table}
The considered cities exhibit striking differences in terms of 
cultural, social, economic, religious and geographic context, and can be roughly divided 
into two large classes: 1) patterns, as Ahmedabad, Cairo and Venice, grown throughout a largely 
self-organized, fine-grained historical process, out of the control of any central agency; 
2) patterns, as Los Angeles, Richmond, and San Francisco, realized over a short period of time as 
the result of a single plan, usually exhibiting a regular grid-like structure. 
The basic characteristics of the derived graphs are reported in Table \ref{table1}.  
$N$ and $K$ assume widely different values, notwithstanding the fact we have considered 
the same amount of land. The edge value (i.e. the street length) distribution, $P(l)$, is single 
peaked in self-organized cities, while it shows many peaks in single planned cities, due 
to their grid patterns \cite{next}.

%
For each of the eighteen cities we have evaluated the four following  
node centrality indices.  

1) {\em Closeness centrality}, $C^C$, measures to which extent a
node $i$ is near to all the other nodes along the shortest
paths, and is defined as  \cite{wasserman94}:  
\begin{equation} 
\label{CC}
C^C_i  = \frac   {N-1}       {  \sum_{ j \in G; j \neq i}   d_{ij}   }
\end{equation}
where  $d_{ij}$ is the shortest path length between $i$ and $j$, defined, 
in a valued graph, as the smallest sum of the edges length $l$  
throughout all the possible paths in the graph between  $i$ and $j$.

2) {\em Betweenness centrality}, $C^B$, is based on the idea that a
node is central if it lies between many other nodes, in the 
sense that it is traversed by many of the shortest paths connecting couples 
of nodes. The betweenness centrality of node $i$ is \cite{wasserman94,freeman}: 
\begin{equation} 
\label{BC}
  C^B_i  = \frac{1}{(N-1)(N-2)} 
{\sum_{{j, k\in G, j \neq k \neq i}}}    n_{jk}(i)/ n_{jk}  
\end{equation}
where $n_{jk}$ is the number of shortest paths between $j$ and $k$, 
and $n_{jk} (i)$ is the number of shortest paths between $j$ and
$k$ that contain node $i$.

3) {\em Straightness centrality}, $C^S$, originates from the idea that
the efficiency in the communication between two nodes $i$
and $j$ is equal to the inverse of the shortest path lenght $d_{ij}$ 
\cite{lm01}.
The straightness centrality of node $i$ is defined as:
\begin{equation} 
\label{SC}
  C^S_i  =  \frac{1}{N-1} { {\sum_{{j \in G, j \neq i}}}  d^{Eucl}_{ij}/ d_{ij} } 
\end{equation}
where $d^{Eucl}_{ij}$ is the Euclidean distance between nodes $i$
and $j$ along a straight line, and we have adopted a normalization recently proposed for 
geographic networks \cite{vragovic}.  
This measure captures to which extent the connecting route between nodes $i$ and $j$ deviates 
from the virtual straight route. 

4) {\em Information centrality}, $C^I$, relates the node centrality  
to the ability of the network to respond to the deactivation of the node 
\cite{centrality}. 
The information centrality of node $i$ is defined 
as the relative drop in the network efficiency  $E[ G]$ caused by the removal 
from $G$ of the edges incident in $i$:
\begin{equation} 
\label{IC}
C^I_i  =  \frac{\Delta E}{E} = 
                \frac{E[G] - E[ G^{\prime}]}{E[ G]}
\end{equation}
where the efficiency of a graph $G$ is defined as \cite{lm01}: 
\begin{equation} 
\label{efficiency}
  E[ G]  =  \frac{1}{N(N-1)} { {\sum_{{i, j \in G, i \neq j }}}  d^{Eucl}_{ij}/ d_{ij}  } 
\end{equation}
and where $G^{\prime}$  is the graph with $N$ nodes and $K-k_i$ edges
obtained by removing from the original graph $G$ the $k_i$ edges incident in node
$i$. An advantange of using the efficiency to measure the performance of a 
graph is that $E[G]$ is finite even for disconnected graphs.

\begin{figure}[htb]
\includegraphics[width=8.cm]{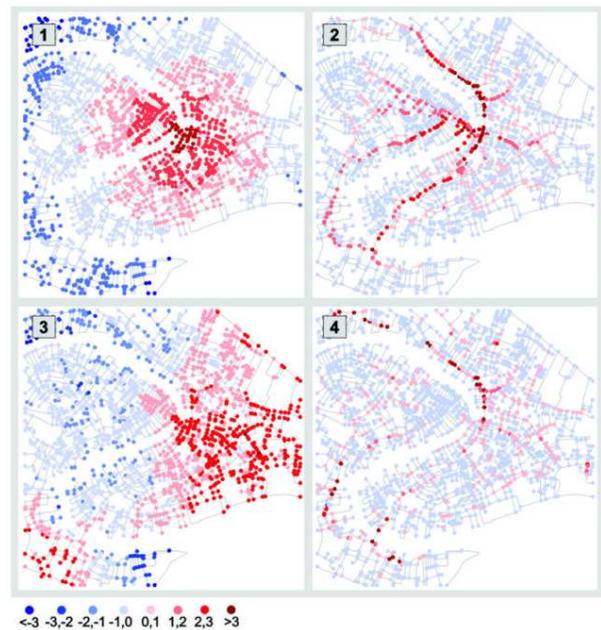}   
\caption{\label{venice} 
Colour-coded maps representing the spatial distributions of node centrality in Venice. 
The four indices:  
(1) closeness $C^C$, (2) betweenness $C^B$, (3) straightness $C^S$ and (4) information $C^I$, 
are visually compared over the spatial graph.  
Different colours represent classes of nodes with different values of centrality (  
the classes are defined in terms of multiples of standard deviations 
from the average, as reported in the colour legend).}
\end{figure}

\begin{figure}[htb]
\includegraphics[width=8.cm]{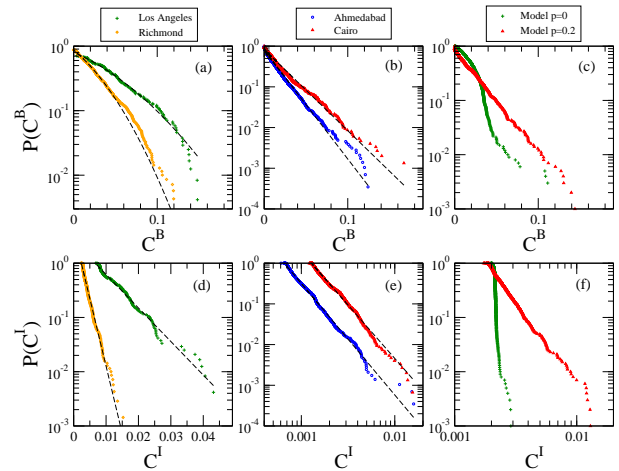}   
\caption{\label{pc} 
Cumulative distributions of betweenness $C^B$ (a, b, c) 
and information $C^I$ (d, e, f) for two planned (Los Angeles and Richmond), 
two self-organized cities (Ahmedabad and Cairo), and the model discussed in the text.  
The cumulative distributions $P(C)$ are defined by 
$P(C) = \int_C^{+\infty} \frac{N(C^{\prime})}{N} dC^{\prime}$, 
where $N(C)$ is the number of nodes having centrality equal to $C$. 
$P(C^B)$ are single scale in all the cases: the dashed lines in panels (a) and (b) are 
respectively exponential, 
$P(C) \sim exp(-C/s)$  ($s_{Ahm} = 0.016$, $s_{Cai}=0.022$), 
and gaussian, $P(C) \sim exp(-x^2/ 2\sigma^2)$ ($\sigma_{LA}= 0.078$, $\sigma_{Rich}=0.049$), 
fits to the empirical distributions. Conversely, $P(C^I)$ 
differentiate self-organized cities 
from planned ones: the dashed lines in the log-log plot of panel (d) indicate that 
the information centrality follows a power law  $P(C) \sim C^{-\gamma}$ for 
the two self-organized 
cities ($\gamma_{Ahm}=2.74$, $\gamma_{Cai}=2.63$), whereas the dashed lines in panel (e) 
indicate an exponential distribution $P(C) \sim exp(-C/s)$ for the two planned cities 
($s_{LA} = 0.007$, $s_{Rich}=0.002$). In panel (f), $P(C^I)$ is exponential in the model 
with $p=0$, and power-law for $p=0.2$. 
}
\end{figure}

%
The spatial distribution of node centralities can be visualized by means 
of colour-coded maps as the one of Venice reported in Fig.~\ref{venice}. 
The figures for the other cities can be downloaded from our website \cite{website}. 
As shown in figure, $C^C$ exhibits a strong trend to group higher scores at the center of the image.  
This is both due to the nature of the index and to the artificial boundaries 
imposed by the 1-sq.~mile maps representation. 
The spatial distribution of $C^B$ nicely captures the continuity of prominent
urban routes across a number of intersections, changes in direction and focal 
urban spots. In the case of Venice the most popular walking
paths and squares (``campi''), and the Rialto bridge over
the Canal Grande, emerge along the red nodes routes. 
In most of the cities considered, $C^B$ is also able to identify  the primary structure of 
movement channels as different to that of secondary, local routes \cite{website}.  
The spatial distribution of $C^S$ depicts both linear routes and focal areas in the urban system:  
$C^S$ takes high values along the main axes, even higher at their intersections.
Finally $C^I$ exhibits a spatial distribution 
that is in many cases similar to that of $C^B$. 
Notwithstanding the similarities in the colour maps, $C^I$ and $C^B$ 
exhibit different statistical distributions, as  illustrated in  Fig.~\ref{pc}, 
where we report an example of the cumulative distributions for the two categories of cities. 
Closeness, straightness (not shown in figure) and betweenness distributions are quite 
similar in both self-organized and planned cities, despite the diversity of
the two cases in socio-cultural and economic terms could not be deeper. 
In particular, $C^B$ exhibits a single scale 
distribution \cite{amaral00} in self-organized and in planned cities, the
former having an exponential distribution, the latter having 
a Gaussian distribution, as respectively shown in Fig.~\ref{pc}a and b. 
Conversely, the distribution of $C^I$ is single-scale for planned cities and 
broad-scale for self-organized cities: e.g. Los Angeles and Richmond are well fitted by 
exponential curves (Fig.~\ref{pc}d), while Ahmedabad and Cairo are fitted by power-laws 
$P(C) \sim C^{- \gamma}$  with exponents $\gamma_{Ahm}=2.74$, $\gamma_{Cai}=2.63$ 
(Fig.~\ref{pc}e). 
Among the self-organized cities, Venice is the one with the
smallest value of the exponent, namely $\gamma_{Ven}=1.49$. This is 
due to the particular environmental constraints that have shaped the historical 
structure of the city. 
The identified power-laws indicate a highly uneven distribution of $C^I$ 
over networks of self-organized cities. This can be considered as the analogous, for 
spatial networks of urban streets, 
of the power-laws observed in the degree and in the betweenness 
distributions of many non-spatial complex networks from biology 
and technology \cite{report}. 
%
%
%
To reproduce the empirical distributions we have considered 
the following model: $N$ nodes are initially placed on a rectangular grid; 
with a probability $p$ each node is moved to
a random position in the unit square; 
for each node $i$, two new edges $(i,j)$ and
$(i,k)$ are added, where $j$ and $k$ are the two nearest nodes among
those not yet connected to $i$. 
The model interpolates from a regular grid, for $p=0$, 
to a graph with randomly placed nodes, for $p=1$. 
The distribution of $C^B$ in the model is single scale for any value of $p$. 
In particular, for values of $p$ in the range $0.1-0.3$, $P(C^B)$ 
is exponential as in self-organized cities. 
Conversely, $P(C^B)$ is single scale for low values of $p$, and follows a 
power law for intermediate values of $p$. 
We have found that the centrality distribution in planned cities are well 
reproduced by the model with $p \sim 0$ (or by triangular and square grids), 
while self-organized cities by the model with $p \sim 0.1-0.3$. 
The distributions obtained for $N=900$, $p=0$ and $p=0.2$,  
are reported in Fig.~\ref{pc}c and ~\ref{pc}f.

%
\begin{figure}[htb]
\includegraphics[width=8.cm]{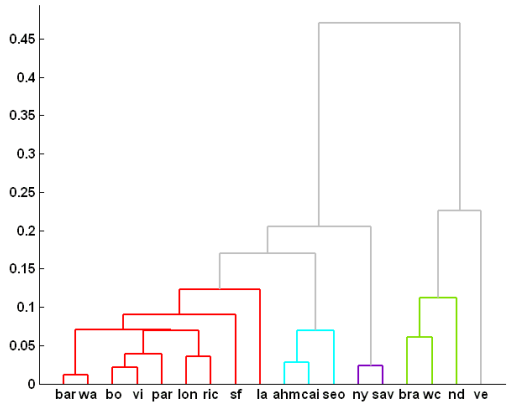}   
\caption{\label{tree} 
Hierarchical tree based on $g^C$, $g^B$, $g^S$ and $g^I$. 
The complete linkage method, based on the largest distance 
between objects in different clusters, has been applied. By choosing a maximum 
distance equal to 0.15 for two cities to belong to the same cluster, 
we find: a first cluster (red) from Barcelona to Los Angeles,  
a second cluster (cyan) from Ahmedabad to Seoul including 
self-organized cities, a third cluster (blue) made up by New York and Savannah, 
a forth cluster (green) from Brasilia to New Delhi, and 
a fifth cluster (grey) constituted only by Venice.}
\end{figure}
Inequalities in the distribution of centrality among the nodes of the network 
can be quantified by evaluating the Gini coefficient of the distribution. 
The Gini coefficient, $g$, is an index commonly 
adopted to measure inequalities of a given resource among the individuals 
of a population. It can be calculated by comparing 
the Lorenz curve of the ranked distribution, i.e. a curve that shows, for the bottom 
$x \%$ of individuals, the percentage $y \%$ of the total resource which they have, 
with the line of perfect equality \cite{dagum}. 
The coefficient $g$ ranges from a minimum value of zero, when all 
individuals are equal, to a maximum value of 1, in a population in which every 
individual, except one, has a size of zero.  
For each of the cities we have evaluated four Gini coefficients, $g^C$, $g^B$, $g^S$, $g^I$, 
one for each of the centrality measures. 
E.g., the Gini coefficient $g^I$ is 0.12 for New York, 0.19 for Richmond, and 0.23 
for Cairo, thus indicating that Cairo has a more heterogeneous information 
centrality distribution than that of Richmond and New York. 
In fig.~\ref{tree} we show the results of a hierarchical 
clustering analysis based on the homogeneity/heterogeneity 
of the networks, as measured by the four Gini coefficients. 
The iterative pairing of cities obtained captures some basic classes of 
urban patterns: it is the case of the early association of Barcelona and Washington 
or New York and Savannah, all grid-iron planned cities as well as that 
of Bologna, Wien and Paris, all mostly 
medieval organic patterns, or that of Ahmedabad and Cairo.  
Brasilia, Walnut Creek and New Delhi, to this respect, share a planned, large 
scale modernist formation. Venice is the last association, 
which tells of the unique mix of fine 
grained pattern and natural constrains that have shaped the historical structure of the city.

We have proposed a comparative analysis of different centrality 
measures in spatial networks of urban streets. 
Each centrality captures a different aspect of one place's 
``being central'' in geographic space, and by the use of many centrality 
measures it is possibile to capture 
structural similarities and dissimilarites across cities. 
Our work opens up to the in-depth investigation of the correlation 
between the structural properties of the system and the relevant dynamics 
on the system, like pedestrian/vehicular flows and 
retail commerce vitality, 
all information traditionally associated to spatial graphs.  
We expect that some of these factors are more strictly correlated to some 
centrality indices than to others, thus giving informed indications 
for strategies of urban planning and design.


\small
\begin{thebibliography}{99}

\bibitem{report} R. Albert and  A.-L. Barab\'asi,  {\it Rev. Mod. Phys.\/} {\bf 74}, 47 (2002); 
M.E.J. Newman, SIAM Review {\bf 45}, 167 (2003).
\bibitem{yook02} S.-H. Yook, H.~Jeong, and A.-L. Barab\'asi, Proc. Natl. Acad. Sci. U.S.A. {\bf 99}, 13382 (2002).
\bibitem{ants04} J. Buhl, J. Gautrais, R.V. Sol\'e, P. Kuntz, S. Valverde, J.L. Deneubourg, and G. Theraulaz, Eur. Phys. J. {\bf B42}, 123 (2004).
\bibitem{newman04} M. T. Gastner and  M. E. J. Newman, cond-mat/0407680.  
\bibitem{west95} D.B. West, \textit{Introduction to Graph Theory}, (Prentice Hall, 1995).
\bibitem{bavelas48} A. Bavelas, \textit{Human Organization} 7, 16 (1948). 
\bibitem{wasserman94} S. Wasserman and K. Faust, {\it Social Networks Analysis}, (Cambridge University Press, Cambridge, 1994).
\bibitem{wilson} G.A. Wilson, {\it Complex Spatial Systems: The Modelling Foundations of Urban and Regional Analysis}, (Prentice Hall, Upper Saddle River, NJ, 2000).
\bibitem{hillier1}   B. Hillier, and J. Hanson, {\it The social logic of space}, (Cambridge University Press, Cambridge, UK, 1984).
\bibitem{jang}       B. Jiang and C. Claramunt, Enviromental and Planning {\bf B31}, 151 (2004). 
\bibitem{porta_epb1} S. Porta, P. Crucitti and V. Latora, Preprint cond-mat/0411241
\bibitem{rosvall}    M. Rosvall, A. Trusina, P. Minnhagen, and K. Sneppen, Phys. Rev. Lett. {\bf 94}, 028701 (2005). 
\bibitem{jacobs} A. Jacobs, {\it Great streets}, (MIT Press, Boston, MA, 1993).
\bibitem{next} P. Crucitti, V. Latora and S. Porta, to be submitted. 
\bibitem{freeman} L.C. Freeman, Social Networks {\bf 1}, 215 (1979). 
\bibitem{lm01} V. Latora and M. Marchiori, Phys. Rev. Lett. {\bf 87}, 198701 (2001). 
\bibitem{vragovic} I. Vragov\`ic, E. Louis and A. D\`\i az-Guilera, Phys. Rev. {\bf E71}, 036122 (2005). 
\bibitem{centrality} V. Latora and M. Marchiori, Preprint cond-mat/0402050; 
Phys. Rev. {\bf E71}, 015103(R) (2005).
\bibitem{website} Website at: http://www.ct.infn.it/~latora/1sqml.html
\bibitem{amaral00} L.A.N.~Amaral, A.~Scala, M.~Barth\'elemy, and H.E.~Stanley, Proc. Natl. Acad. Sci. (USA) {\bf 97}, 11149 (2000).
\bibitem{dagum} C. Dagum, Econ. Appl. {\bf 33}, 327 (1980). 






\end{thebibliography}
\end{document}